\documentclass[aps,pre,graphicx,reprint,superscriptaddress]{revtex4-1}

\usepackage{amsmath}
\usepackage{amsfonts}
\usepackage{graphicx}
\usepackage{amssymb}
\usepackage{bm}
\usepackage{braket}
\usepackage{hyperref}
\usepackage{tikz}

\begin{document}

\title{Approach to nonlinear magnetohydrodynamic simulations in stellarator geometry}

\author{Yao Zhou}
\email[]{yaozhou@princeton.edu}
\affiliation{Princeton Plasma Physics Laboratory, Princeton, New Jersey 08543, USA}
\author{N.~M.~Ferraro}
\affiliation{Princeton Plasma Physics Laboratory, Princeton, New Jersey 08543, USA}
\author{S.~C.~Jardin}
\affiliation{Princeton Plasma Physics Laboratory, Princeton, New Jersey 08543, USA}
\author{H.~R.~Strauss}
\affiliation{HRS Fusion, West Orange, New Jersey 07052, USA}

\date{\today}

\begin{abstract}
The capability to model the nonlinear magnetohydrodynamic (MHD) evolution of stellarator plasmas is developed by extending the M3D-$C^1$ code to allow non-axisymmetric domain geometry. We introduce a set of logical coordinates, in which the computational domain is axisymmetric, to utilize the existing finite-element framework of M3D-$C^1$. A $C^1$ coordinate mapping connects the logical domain to the non-axisymmetric physical domain, where we use the M3D-$C^1$ extended MHD models essentially without modifications. We present several numerical verifications on the implementation of this approach, including simulations of the heating, destabilization, and equilibration of a stellarator plasma with strongly anisotropic thermal conductivity, and of the relaxation of stellarator equilibria to integrable and non-integrable magnetic field configurations in realistic geometries.

\end{abstract}

\maketitle

\section{Introduction}
A major advantage of the stellarator concept over the tokamak is its superior magnetohydrodynamic (MHD) stability \cite{Shafranov1983}. Not requiring plasma currents to generate the confining magnetic fields, stellarators are generally free of current-driven instabilities that can be disruptive in tokamaks. Still, stellarator plasmas can be subject to pressure-driven instabilities, and designs usually rely on linear stability analysis to avoid them, which in turn imposes limits on the theoretically achievable plasma beta. 

However, stellarator plasmas are often observed to be nonlinearly stable when driven beyond linear stability thresholds in experiments \cite{Weller2006}. Linearly unstable modes are seen to grow but typically saturate at harmlessly low levels, implying that linear stability constraints tend to be overly conservative and restrictive. Hence, it would be useful to consider nonlinear stability criteria instead, which may expand operation windows for present devices and improve designs and lower costs for future ones.

\textcolor{black}{Unfortunately, systematic investigations on this idea have been impeded by the lack of a state-of-the-art nonlinear MHD code for stellarators.} Most existing toroidal MHD codes are designed for tokamak applications and therefore assume axisymmetric computational domains. While some simple stellarators can be modeled using such codes \cite{Schlutt2012,Schlutt2013,Roberds2016}, most realistic stellarator designs do not permit an axisymmetric surface between the plasma and the coils, and therefore cannot be treated using axisymmetric domains. To our knowledge, the M3D \cite{Strauss2004} and MIPS \cite{Sato2017} codes have developed the capability to allow non-axisymmetric domains, but they have not been used for simulations at transport timescale. \textcolor{black}{Lately, NIMROD \cite{Sovinec2020} and JOREK \cite{Nikulsin2021} have also been exploring this possibility.}

In this work, we fill this need by extending the M3D-$C^1$ code \cite{Jardin2012} from tokamak to stellarator geometry. For time advance, M3D-$C^1$ implements a split-implicit scheme that allows for time steps larger than Alfv\'enic, which realizes stable transport-timescale simulations \cite{Jardin2012b}. For 3D spatial discretization, M3D-$C^1$ uses high-order finite elements with $C^1$ continuity (see Section \ref{approach} for details), which are constructed on an axisymmetric mesh. To utilize this finite-element framework, we introduce a set of logical coordinates, in which the computational domain becomes axisymmetric. A $C^1$ mapping connects the logical coordinates to the physical $(R, Z, \varphi)$ coordinates so that we can use the chain rule to calculate derivatives in the latter, in terms of which the existing physics equations are written. This way, we can readily use the MHD models within M3D-$C^1$ without introducing new metric factors or coordinate singularities, and the physics coding carries over essentially without modification. 

We present results from several numerical tests to verify the implementation of this approach. First is a convergence study on a boundary-value problem, solving Laplace's equation in a stellarator-shaped domain. Then, we perform nonlinear MHD simulations of the heating of a rotating-ellipse stellarator, which are done in either a non-axisymmetric or axisymmetric domain, and compare the results for benchmarking. Finally, we demonstrate the capability to treat realistic geometries by studying the relaxation of VMEC \cite{Hirshman1983} equilibria, including cases where flux surfaces generally stay intact or break up due to pressure-driven currents.

This paper is organized as follows. In Section \ref{approach}, we describe our approach to extending M3D-$C^1$ to stellarator geometry. In Section \ref{verification}, we present numerical results to verify the implementation of this approach. Summary and discussion follow in Section \ref{summary}.

\section{Approach to stellarator geometry}\label{approach}
Let us first review how M3D-$C^1$ treats 3D tokamak geometry, where cylindrical coordinates $(R,Z,\varphi)$ are used. [The order of coordinates does not imply handedness and is merely chosen for convenience; M3D-$C^1$ actually uses a right-handed $(R,\varphi,Z)$ coordinate system.] An axisymmetric domain is discretized using wedge-shaped $C^1$ elements, which are tensor products of reduced quintic triangular elements \cite{Jardin2004} in the $(R,Z)$ plane and Hermite cubic elements \cite{Strang1973} in the toroidal $(\varphi)$ direction. In such an `extruded' element, a scalar function $g(R,Z,\varphi)$ can be projected onto basis functions $\nu(R,Z,\varphi)$:
\begin{gather}
g(R,Z,\varphi)=\sum_{j=1}^{18}\sum_{k=1}^{4}g_{jk}\nu_{jk}(R,Z,\varphi)\label{basisRZ},
\end{gather}
where $\nu_{jk}(R,Z,\varphi)=\xi_{j}(R,Z)h_k(\varphi)$ and $\xi(R,Z)$ and $h(\varphi)$ denote the basis functions of reduced quintic and Hermite cubic elements, respectively. The degrees of freedom (DoFs) $g_{jk}$ are given by $(g,g_R,g_Z,g_{RR},g_{RZ},g_{ZZ})$ and their $\varphi$ derivatives
on all six nodes of the element. (Coordinates in subscripts denote partial derivatives.) In the Galerkin method, derivatives up to second order are allowed on these $C^1$-continuous basis functions $\nu$ (up to fourth order considering integration by parts) \cite{Strang1973}. 

\begin{figure}
\includegraphics[scale=0.45]{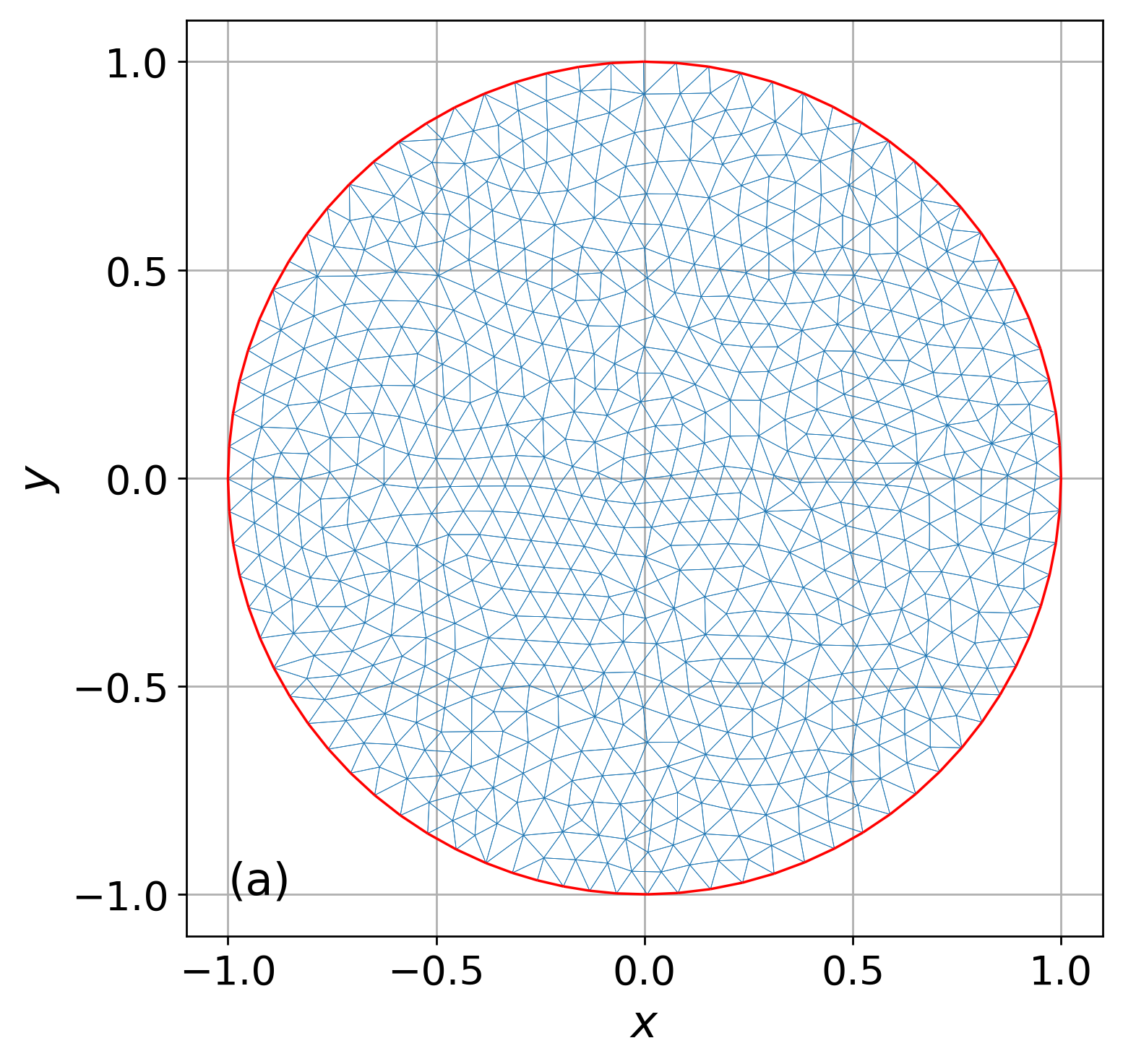}
\includegraphics[scale=0.45]{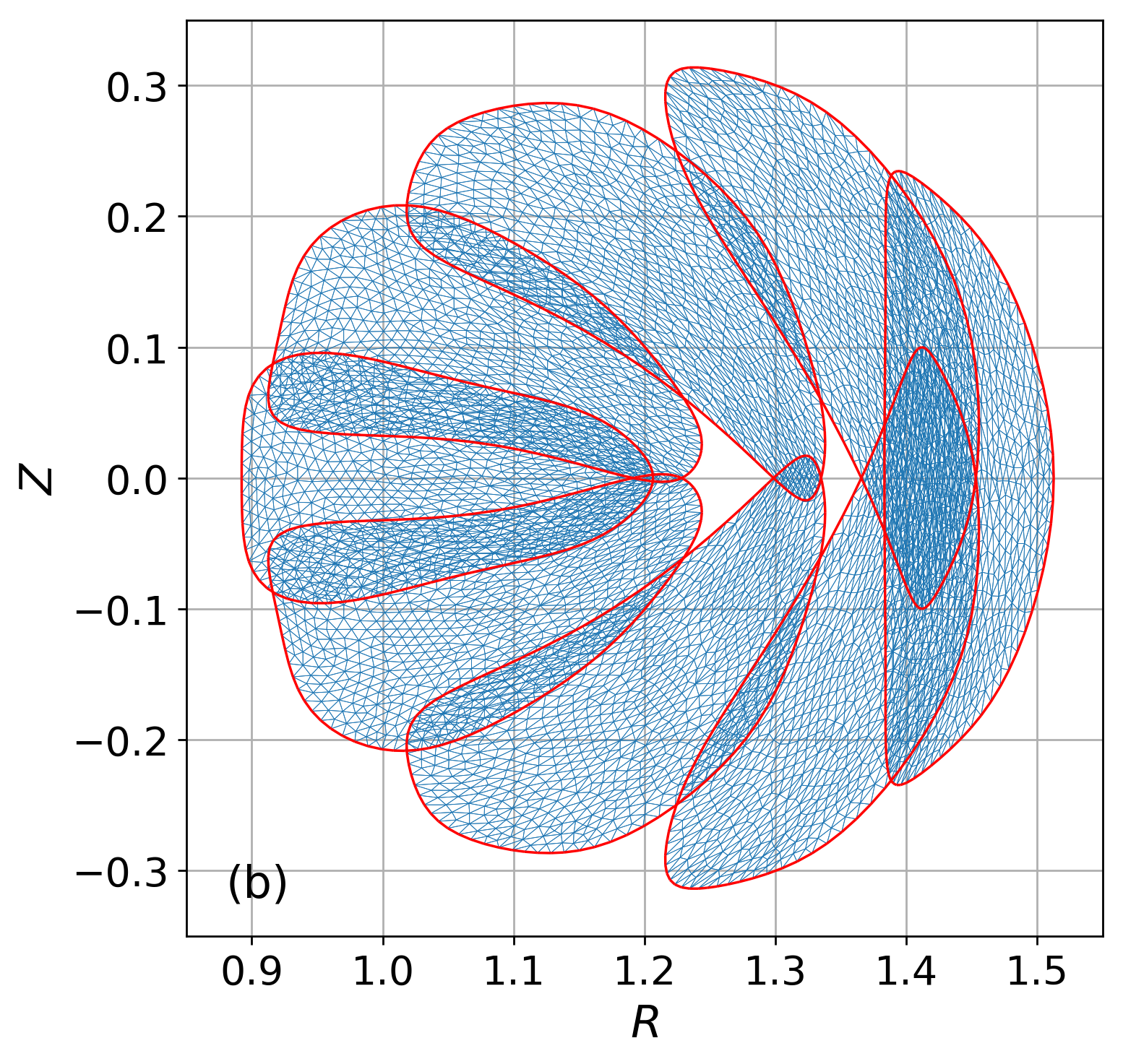}
\caption{\label{meshfig}Toroidal cross sections of (a) an axisymmetric mesh in the logical coordinates, and (b) the non-axisymmetric mesh that it maps to in the physical coordinates, at eight different toroidal angles. The mapping is generated using an HSX-like VMEC equilibrium.}
\end{figure}

The M3D-$C^1$ finite elements described above must be constructed on an axisymmetric mesh, which is natural for tokamak simulations. However, the capability to treat non-axisymmetric computational domains is essential for modeling realistic stellarators with complex geometries. To facilitate this, we introduce a set of logical coordinates $(x,y,\zeta)$, in which the domain is axisymmetric. These coordinates connect to the physical $(R,Z,\varphi)$ coordinates via a $C^1$-diffeomorphic mapping
\begin{gather}
R = R(x,y,\zeta),~Z = Z(x,y,\zeta),~ \varphi = \zeta.\label{xy}
\end{gather}
While it is possible to consider more general mappings between the logical and physical toroidal angles, here we simply equate $\zeta$ and $\varphi$ for practicality. In principle, the mapping \eqref{xy} does not need to have any physical meanings, but a particularly convenient choice is to utilize the outputs of equilibrium codes like VMEC \cite{Hirshman1983}, which are given in terms of the geometries of nested flux surfaces, $R(s,\theta,\zeta)$ and $Z(s,\theta,\zeta)$. With $s$ being a surface label and $\theta$ being a poloidal angle, we can use a polar--Cartesian transformation, $x=\sqrt{s}\cos\theta$ and $y=\sqrt{s}\sin\theta$, to obtain the logical coordinates. Figure \ref{meshfig} shows such a mapping generated using an HSX-like VMEC equilibrium \cite{Talmadge1996}.

Since the computational domain is axisymmetric in the logical $(x,y,\zeta)$ coordinates, we can readily utilize the existing M3D-$C^1$ finite-element framework for spatial discretization by using $(x,y,\zeta)$ in lieu of $(R,Z,\varphi)$, as is shown in Figure \ref{meshfig}(a). In this case, a scalar function $g$ is now projected onto logical basis functions $\nu(x,y,\zeta)$,
\begin{gather}
g(x,y,\zeta)=\sum_{j=1}^{18}\sum_{k=1}^{4}g_{jk}\nu_{jk}(x,y,\zeta)\label{basisxy},
\end{gather}
where $\nu_{jk}(x,y,\zeta)=\xi_{j}(x,y)h_k(\zeta)$ and the DoFs $g_{jk}$ are given by $(g,g_x,g_y,g_{xx},g_{xy},g_{yy})$ and their $\zeta$ derivatives on the nodes. While these basis functions allow derivatives up to second order in $(x,y,\zeta)$, the existing physics equations are written in terms of their derivatives with respect to the physical $(R,Z,\varphi)$ coordinates, which we can obtain using the chain rule. Specifically, first-order physical derivatives are given by
\begin{align}\label{xy2RZ}
\begin{pmatrix} \nu_R \\ \nu_Z\\ \nu_\varphi \end{pmatrix}
=J\begin{pmatrix} \nu_x \\ \nu_y \\ \nu_\zeta \end{pmatrix},
\end{align}
where the Jacobian matrix $J$ is
\begin{align}
J
= \begin{pmatrix} R_x & Z_x & 0 \\ R_y & Z_y & 0 \\ R_\zeta & Z_\zeta & 1 \end{pmatrix}^{-1}
=\frac{1}{D}\begin{pmatrix} Z_y & -Z_x & 0 \\ -R_y & R_x & 0 \\ A & B & D \end{pmatrix},
\end{align}
with
\begin{subequations}\label{ABD}
\begin{gather}
A = R_y Z_\zeta-R_\zeta Z_y,\\
 B = R_\zeta Z_x-R_x Z_\zeta,\\
  D = R_x Z_y-R_y Z_x.
\end{gather}
\end{subequations}
The transformations of second derivatives are much more cumbersome, which we summarize in Appendix \ref{second}. Note that the discrete equations are derived using the Galerkin method assuming that first derivatives $(\nu_R,\nu_Z,\nu_\varphi)$ are continuous. This means that the coordinate mapping \eqref{xy} must be $C^1$-diffeomorphic, which we \textcolor{black}{choose to} guarantee by representing it with our $C^1$-continuous basis functions: $R(x,y,\zeta)=\sum R_{jk}\nu_{jk}$, $Z(x,y,\zeta)=\sum Z_{jk}\nu_{jk}$.

Meanwhile, we also need to keep track of the Jacobian determinant when performing volume integrals:
\begin{gather}
\int g\,\mathrm{d}V = \int g\,R\,\mathrm{d}R\,\mathrm{d}Z\,\mathrm{d}\varphi\,= \int g\,RD\,\mathrm{d}x\,\mathrm{d}y\,\mathrm{d}\zeta. 
\end{gather}
Another subtlety is that for boundary conditions to be imposed, the DoFs $g_{jk}$ also need to be transformed from logical to physical derivatives. The treatment is discussed in detail in Appendix \ref{boundary}. Notably, all the modifications described above are made on the level of basis functions, such that no significant changes to the extended MHD models implemented in M3D-$C^1$ are required.

Finally, we remark that such a coordinate mapping is a common approach when structured finite elements are used to discretize shaped domains. In fact, NIMROD \cite{Sovinec2004} and JOREK \cite{Czarny2008} already use axisymmetric mappings in tokamak geometry. This was not necessary for M3D-$C^1$ since the wedge-shaped elements can mesh arbitrary axisymmetric domains directly, but stellarator geometry requires the implementation of a non-axisymmetric mapping because the elements are structured in the toroidal direction. Moreover, unlike NIMROD and JOREK, M3D-$C^1$ uses $C^1$ elements and hence requires the transformation of second derivatives, which introduces some complication (c.f.~Appendix \ref{second}). Also, both NIMROD and JOREK use Fourier discretization toroidally, and therefore efforts to adapt these codes to non-axisymmetric domains would presumably somewhat differ from the approach taken here.

\section{Numerical verifications}\label{verification}
\subsection{Boundary-value problem}
First, let us verify the spatial discretization of a non-axisymmetric domain by considering a boundary-value problem. Specifically, we solve Laplace's equation in 3D
\begin{align}\label{Laplace}
\nabla^2\phi = 0,
\end{align}
where $\phi$ is a scalar field. We consider not toroidal but periodic cylindrical geometry, where the analytical solution comprises components
\begin{align}
\phi_{mn} = \epsilon_{mn}I_m(nr/R_0)\cos(m\theta - nz/R_0).\label{Bessel}
\end{align}
Here, $m$ and $n$ are integers denoting poloidal and toroidal (axial) mode numbers, respectively, and $I_m$ denotes modified Bessel function of the first kind. The periodic cylindrical coordinates $(r,\theta,z)$ relate to $(R,Z,\varphi)$ by $R = R_{\text{a}}+r\cos\theta$, $Z = Z_{\text{a}}+r\sin\theta$, and $\varphi = z/R_0$, where $R_0$ is an effective major radius and $(R_{\text{a}},Z_{\text{a}})$ locates the axis of the cylinder.

\begin{figure}
\includegraphics[scale=0.4]{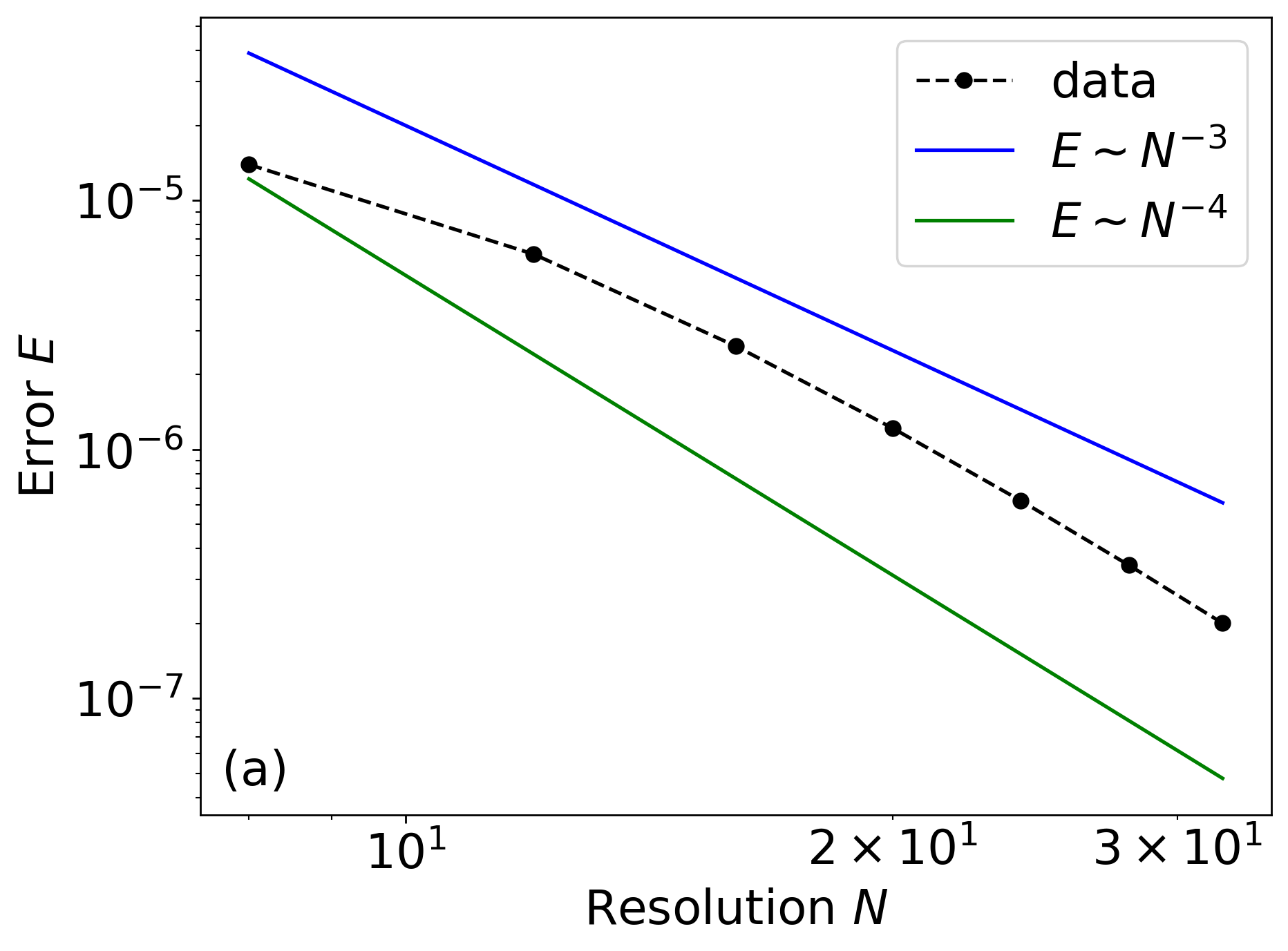}
\includegraphics[scale=0.4]{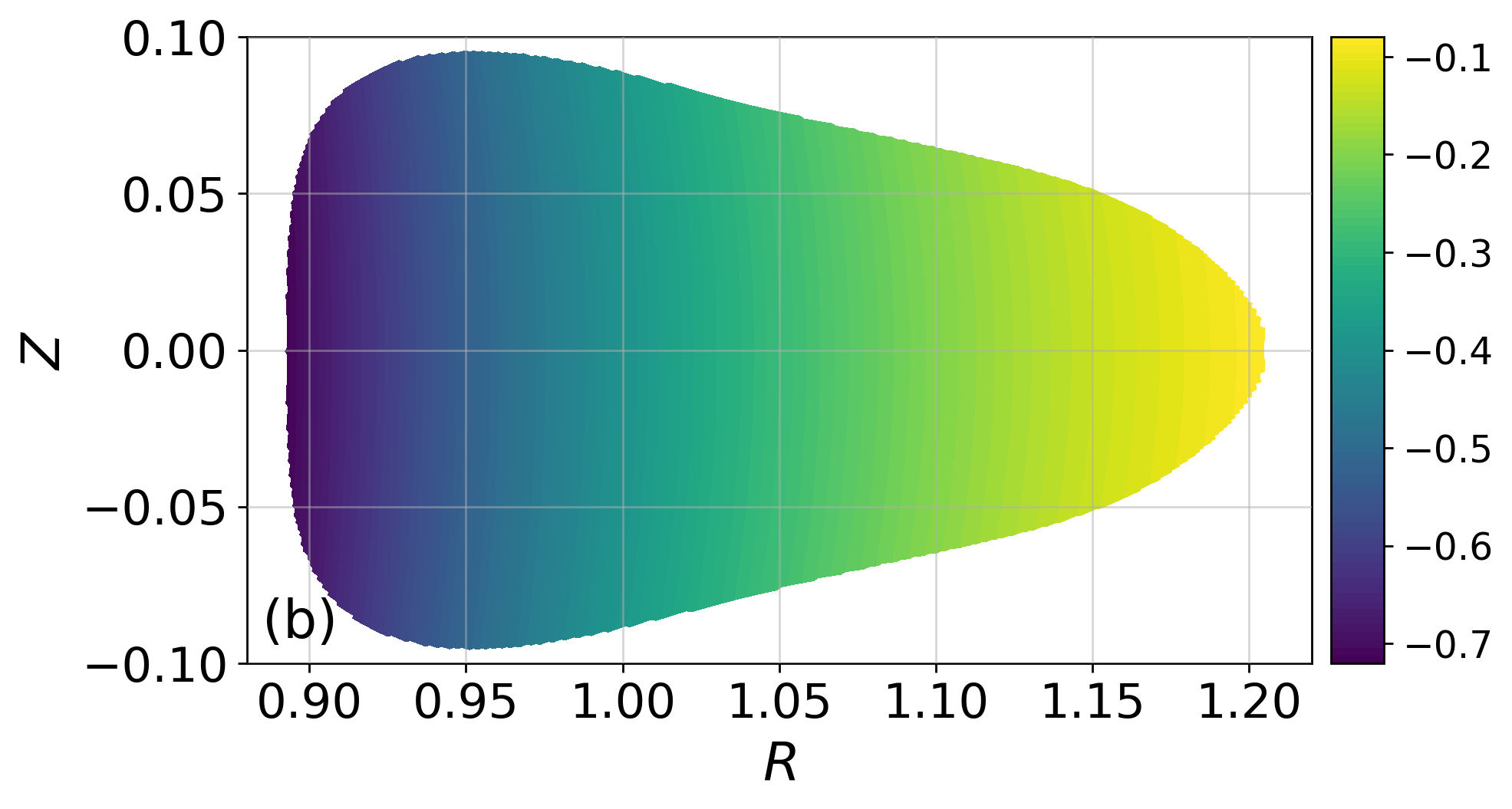}
\caption{\label{convfig}Convergence test on a boundary-value problem in an HSX-shaped domain (only one field period is used): (a) mean-squared error $E$ versus numerical resolution $N$; (b) a cross section of the numerical solution $\phi$ at $\varphi=\pi/4$. The parameters used are $m=2$, $n=1$, $R_0=0.25$, $\epsilon=0.5$, $R_{\text{a}}=1.4$, and $Z_{\text{a}}=0$.}
\end{figure}

In a stellarator-shaped domain as depicted in Figure \ref{meshfig}(b), we impose Dirichlet boundary condition on $\phi$ using the boundary value of the solution \eqref{Bessel}. Then, we obtain numerical solutions to \eqref{Laplace} while increasing the numerical resolution consistently in all three dimensions with the number of toroidal planes $N$ as an indicator. In Figure \ref{convfig}(a), the mean-squared error with respect to the analytical solution shows convergence close to fourth order versus increasing numerical resolution, which verifies our approach to spatially discretizing the non-axisymmetric domain.

\subsection{Dynamical benchmark}
Next, we simulate the heating of a rotating-ellipse stellarator in periodic cylindrical geometry. The simulation can be done in either an axisymmetric or non-axisymmetric computational domain so we can use the original (tokamak) version of M3D-$C^1$ for benchmarking. 

Specifically, we initialize with a vacuum magnetic field $\mathbf{B}=\nabla\phi$ with the potential $\phi$ satisfying \eqref{Laplace}, including an $m=2$, $n=1$ component of solution \eqref{Bessel}, 
\begin{align}
\phi = B_0[z+2\epsilon R_0 I_2(r/R_0)\cos(2\theta - z/R_0)],\label{ellipse}
\end{align}
where $B_0$ denotes the strength of the toroidal (axial) field. This vacuum field generates rotating elliptical flux surfaces near the axis when $\epsilon <1$, and we choose $\epsilon = 0.8$ here, which provides a rotational transform $\iota = 0.2$ and elongation $e = 3$ on the axis \cite{Shafranov1980}. Hence, we can simulate this stellarator in a non-axisymmetric, rotating elliptical domain with lengths of semi-major axis $a=0.3$ and semi-minor axis $b=0.1$. Meanwhile, we can also simulate it in an axisymmetric, cylindrical domain with minor radius $a=0.3$, akin to the NIMROD simulations in \cite{Schlutt2013}. In both cases, the last closed flux surface is limited at $a=0.3$, so the behavior of the plasma, heat transport in particular, should be quite similar, even though exact agreement should not be expected.

While two-fluid and other effects are available in M3D-$C^1$, we solve the single-fluid extended MHD equations in these simulations, including the momentum equation for the fluid velocity $\mathbf{v}$ (in dimensionless units)
\begin{gather}\label{momentum}
\rho(\partial_t \mathbf{v} + \mathbf{v}\cdot\nabla\mathbf{v}) = \mathbf{j}\times\mathbf{B} - \nabla p - \nabla\cdot\mathbf{\Pi},
\end{gather}
the energy equation for the fluid pressure $p$
\begin{align}\label{energy}
\partial_t p + \mathbf{v}\cdot\nabla p +\Gamma p\nabla\cdot\mathbf{v} &= \nonumber\\
(\Gamma-1)(\eta j^2 &- \nabla\cdot\mathbf{q} - \Pi : \nabla\mathbf{v}+Q),
\end{align}
and the induction equation for the magnetic field $\mathbf{B}$
\begin{gather}\label{induction}
\partial_t \mathbf{B} =  \nabla\times(\mathbf{v}\times\mathbf{B} - \eta \mathbf{j}),
\end{gather}
with the current density $ \mathbf{j}$ given by Ampere's law, $ \mathbf{j}= \nabla\times\mathbf{B}  $. Here we do not solve the continuity equation but hold the mass density $\rho=1$ such that \eqref{energy} is essentially a temperature equation. \textcolor{black}{(Otherwise, the agreement would be not as good due to the discrepancy in the density evolution beyond the last closed flux surface, which would undermine the purpose of this benchmark.)} The stress tensor is given by $\mathbf{\Pi} = -\mu(\nabla\mathbf{v}+\nabla\mathbf{v}^{\text{T}}) - 2(\mu_{\text{c}}-\mu)(\nabla\cdot\mathbf{v})\mathbf{I}$ and the heat flux \textcolor{black}{$\mathbf{q} = -\kappa_\perp\nabla T - \kappa_\parallel\mathbf{b}\mathbf{b}\cdot\nabla T$}, with $\mathbf{b} =\mathbf{B}/B$ and the temperature $T=Mp/\rho$, where $M$ is the ion mass. Transport coefficients include resistivity $\eta$, isotropic and compressible viscosities $\mu$ and $\mu_{\text{c}}$, and perpendicular and parallel thermal conductivities $\kappa_\perp$ and $\kappa_\parallel$, and $\Gamma=5/3$ is the adiabatic index. An axisymmetric Gaussian heat source $Q=w/(2\pi\sigma^2)\,\mathrm{e}^{-r^2/(2\sigma^2)}$ is applied in the simulations to heat the plasma, with $w$ and $\sigma$ denoting the heating rate and the Gaussian width, respectively. 

\begin{figure}
\includegraphics[scale=0.4]{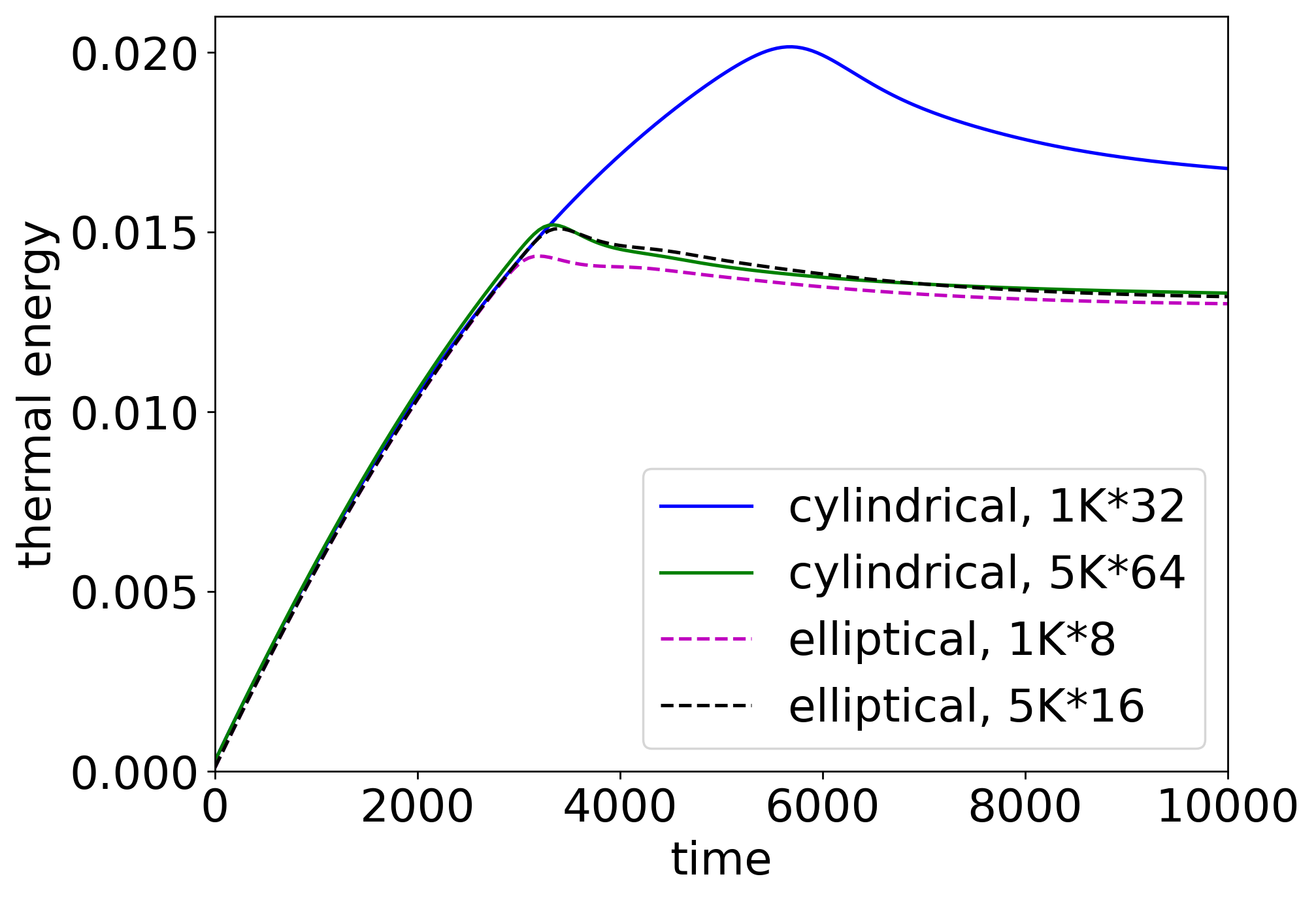}
\caption{\label{history}Total thermal energy versus time in simulations using a cylindrical (solid) or a rotating elliptical (dashed) domain. The number of elements at each toroidal plane is 1389 (1K) in lower resolution runs and 5131 (5K) in higher resolution runs. Simulations in the cylindrical domain require more toroidal planes (32 and 64) than the rotating elliptical domain (8 and 16) for comparable accuracy. The simulation parameters are $B_0=1$, $R_0=1$, $R_{\text{a}}=1.4$, $Z_{\text{a}}=0$, $\epsilon=0.8$, $\sigma=0.05$, $w=\kappa_\perp=\eta=10^{-6}$, $\kappa_\parallel=1$, and $\mu=\mu_{\text{c}}=10^{-4}$. }
\end{figure}

\begin{figure}
\includegraphics[scale=0.4]{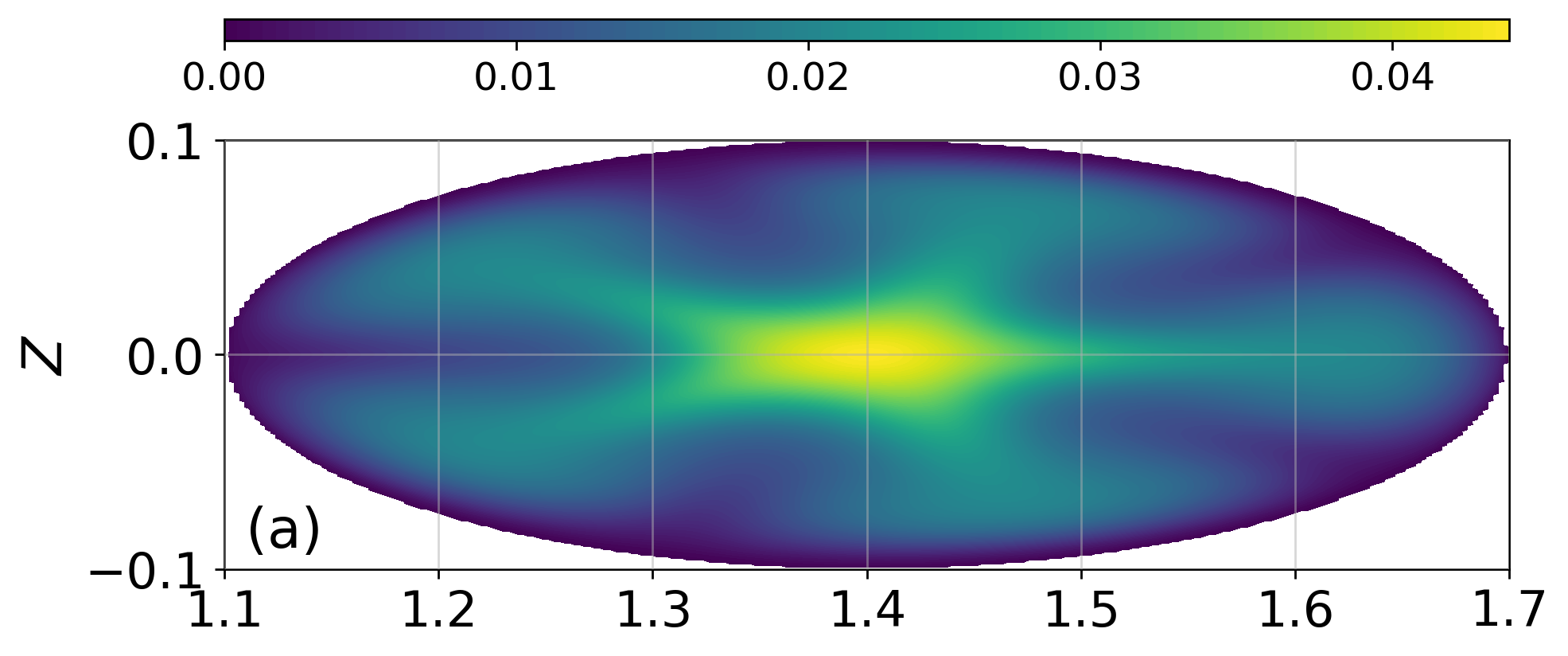}
\includegraphics[scale=0.4]{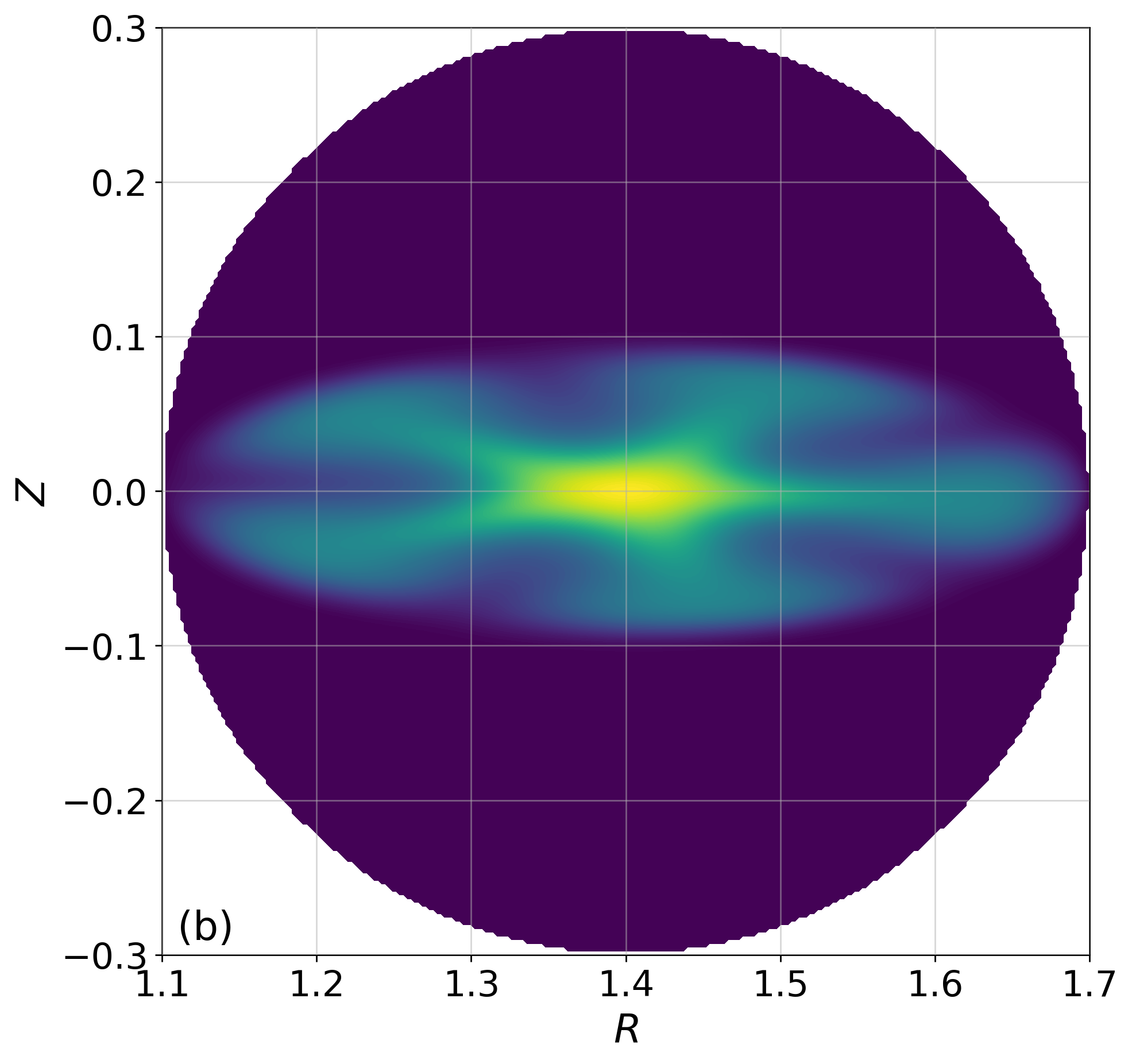}
\caption{\label{interchange}Final snapshot of pressure $p$ at $\varphi=0$ in the higher resolution simulations using (a) the rotating elliptical domain and (b) the cylindrical domain.}
\end{figure}

In Figure \ref{history}, we see an agreement on the growth of thermal energy in the early stage, which verifies that simulations in the non-axisymmetric domain model anisotropic heat transport as accurately as those in the axisymmetric domain. In fact, we find that the axisymmetric domain requires much higher toroidal resolution to produce comparable results, which suggests that the non-axisymmetric domain is more efficient in treating the helical structure. (Similar findings have also been reported in NIMROD simulations \cite{Sovinec2020}.) \textcolor{black}{This is not surprising since the mesh is better aligned with the flux surfaces in the non-axisymmetric domain.} In the later stage, an $m=5$, $n=1$ interchange instability is triggered and the plasma eventually equilibrates. The onsets of the instability depend on the perturbations, which are not prescribed but spontaneous, and hence do not agree exactly. Still, the equilibrium structures in the final equilibria do show qualitative agreement in Figure \ref{interchange}. In summary, the newly implemented \textcolor{black}{stellarator extension can model MHD instabilities and anisotropic heat transport with similar accuracy to the tokamak version but lower computational costs.}

\begin{figure}
\includegraphics[scale=.4]{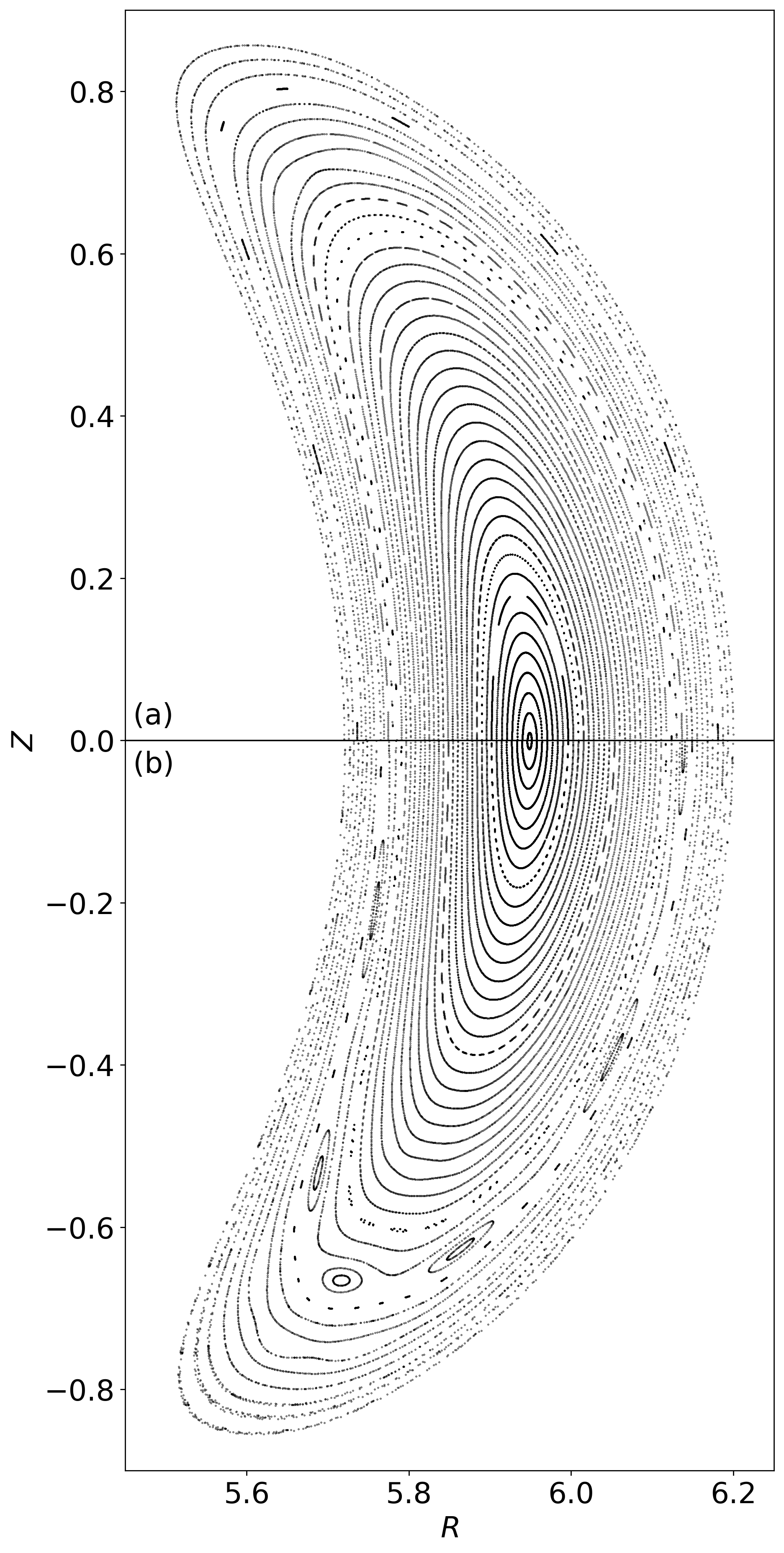}
\caption{\label{w7x}Poincar\'e plots of (a) the initial magnetic field that is interpolated from a W7-X VMEC equilibrium (normalized by Tesla and meter), and (b) the relaxed magnetic field at $t=1000$. The simulation parameters are $\kappa_\perp=\eta=10^{-6}$, $\kappa_\parallel=1$, and $\mu=\mu_{\text{c}}=10^{-4}$. Only one of five field periods is simulated.}
\end{figure}

\subsection{VMEC equilibria}\label{vmec}
Finally, we demonstrate that our approach can treat realistic, strongly shaped stellarator geometries by examining the relaxation of VMEC equilibria. That is, we set up initial conditions in M3D-$C^1$ by interpolating VMEC equilibria and then solve \eqref{momentum}-\eqref{induction} without source terms. Due to the low near-axis resolution in VMEC data, we use Zernike polynomials \cite{Zernike1934} for radial interpolation to ensure smoothness. These polynomials guarantee analyticity near the magnetic axis and have recently been used in stellarator equilibrium codes like SPEC \cite{Qu2020} and DESC \cite{Dudt2020} as well.

\begin{figure}
\includegraphics[scale=.4]{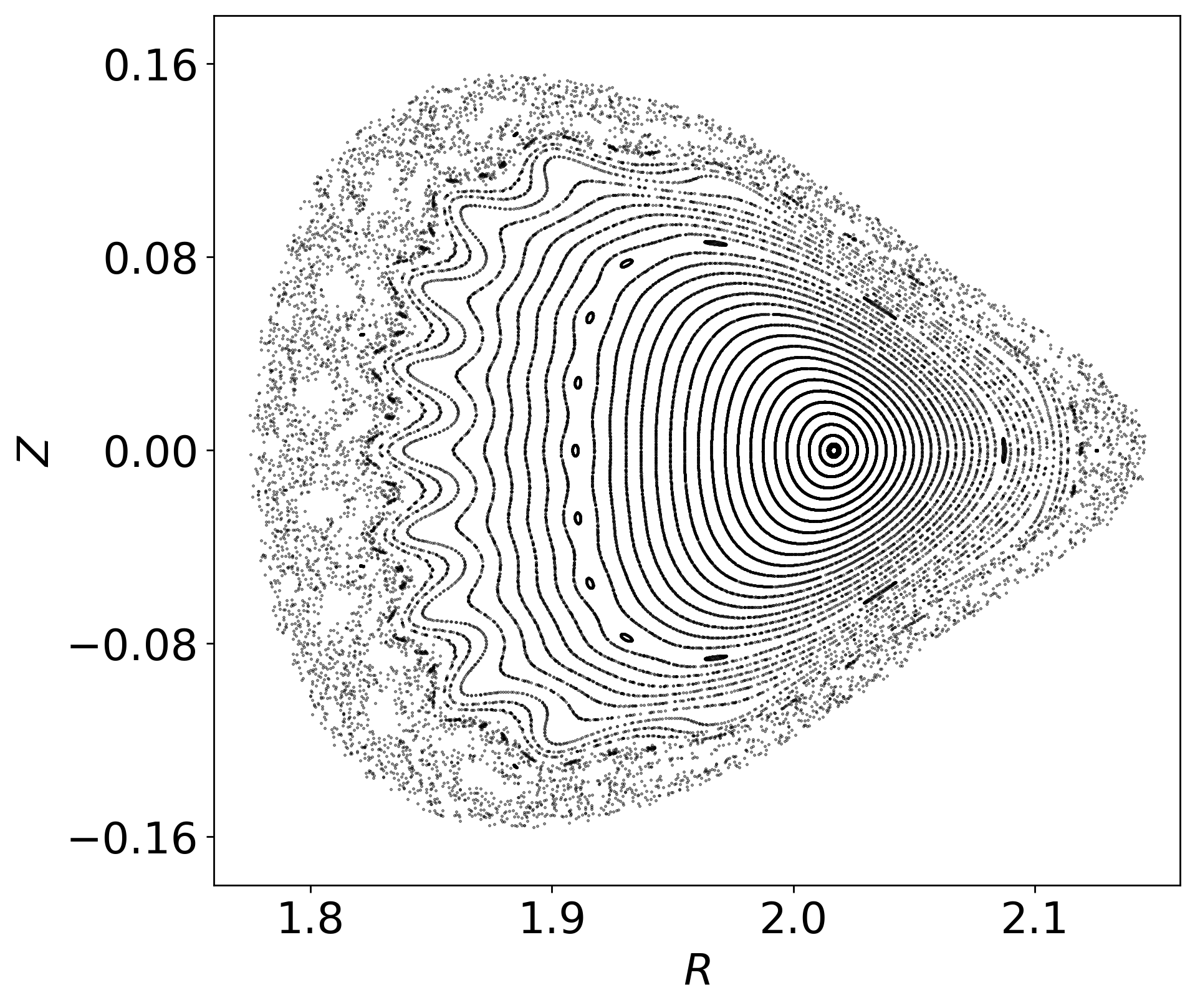}
\caption{\label{w7as}Poincar\'e plot of the magnetic field at $t=100$ in the relaxation of a high-beta W7-AS VMEC equilibrium (normalized by Tesla and meter). The simulation parameters are $\kappa_\perp=\eta=10^{-6}$, $\kappa_\parallel=1$, and $\mu=\mu_{\text{c}}=10^{-3}$. Only one of five field periods is simulated.}
\end{figure}

The first case we study is a W7-X equilibrium with no net toroidal current and no pressure \cite{Beidler1990}, \textcolor{black}{which is close to but not exactly a vacuum field}. Figure \ref{w7x}(a) shows the initial magnetic field constructed in M3D-$C^1$, where flux surfaces are nested as is assumed in VMEC. Figure \ref{w7x}(b) shows the relaxed magnetic field at $t=1000$, where the majority of flux surfaces stay intact and barely displaced. An $m=11$ island chain has emerged at the $\iota = 10/11$ surface, which is not surprising because VMEC solutions are known to be inaccurate at rational surfaces where current singularities can open up islands \cite{Hirshman2011,Helander2014}. Overall, this case exemplifies that M3D-$C^1$ can, to a large extent, sustain a VMEC equilibrium that is supposed to be relatively accurate.

In contrast, the second case we consider is a W7-AS high-beta equilibrium where pressure-induced breaking of flux surfaces has been shown \cite{Zarnstorff2004} using the PIES code \cite{Reiman1986}. Figure \ref{w7as} shows the magnetic field configuration at $t=100$ with a large stochastic region near the edge as well as a pronounced $m=13$ structure. These features are also seen in Figure 7(A) in \cite{Zarnstorff2004}, which was obtained using PIES from the same VMEC equilibrium. Due to the enhanced heat loss by the stochastic magnetic field and the absence of a source,  pressure decreases at a rate comparable to the evolution of the magnetic geometry and a near-equilibrium state could not be reached here. Therefore, a quantitative comparison with PIES results has not been performed and will be left for future work.

\section{Summary and discussion}\label{summary}
In this work, we develop the capability to model the nonlinear MHD evolution of stellarator plasmas by extending the M3D-$C^1$ code to allow non-axisymmetric domain geometry. We introduce a set of logical coordinates, in which the computational domain is axisymmetric, to utilize the existing finite-element framework of M3D-$C^1$. Via the chain rule, the $C^1$ mapping from the logical to the physical $(R, Z, \varphi)$ coordinates facilitates calculations of derivatives in the latter, in terms of which the existing physics equations are written. This way, no significant changes to the extended MHD models within M3D-$C^1$ are required. 

Several numerical verifications on the implementation of this approach are presented. First is a convergence test on a boundary-value problem in stellarator geometry. Then we compare nonlinear simulations of a rotating-ellipse stellarator in a non-axisymmetric domain to those in an axisymmetric domain, and the results show good agreement in terms of the heating, destabilization, and equilibration of the plasma. Finally, we show proof-of-principle simulations of the relaxation of VMEC equilibria to integrable and non-integrable magnetic field configurations, which demonstrate our capability to treat realistic stellarator geometries.

The simulations in section \ref{vmec} are fixed-boundary for including only the plasma region. We can initialize free-boundary simulations once an interface with the vacuum region of free-boundary VMEC \cite{Hirshman1986} is implemented. This could facilitate more rigorous comparisons with more sophisticated equilibrium codes such as PIES \cite{Reiman1986}, HINT \cite{Suzuki2006}, and SPEC \cite{Hudson2012}. We also have plans to verify M3D-$C^1$ against linear stability codes such as TERPSICHORE \cite{Anderson1990} and CAS3D \cite{Nuhrenberg1999}. Furthermore, validation against experimentally observed MHD events such as the sawtooth-like oscillations induced by current drive in W7-X \cite{Zanini2020} or the core collapses in inwardly shifted high-beta LHD configurations \cite{Weller2006} would be of interest as well.

\acknowledgments
We thank A.~H.~Reiman for providing the W7-AS high-beta equilibrium files, and S.~R.~Hudson, D.~A.~Gates, C.~Liu, C.~R.~Sonivec, A.~M.~Wright, and C.~Zhu for helpful discussions. This work was supported by the U.S. Department of Energy under contract number DE-AC02-09CH11466. The United States Government retains a non-exclusive, paid-up, irrevocable, world-wide license to publish or reproduce the published form of this manuscript, or allow others to do so, for United States Government purposes.

\appendix
\section{Transformations of second derivatives}\label{second}
Here, we summarize the expressions of second-order physical derivatives. Those involving $(R,Z)$ only are
\begin{subequations}\label{RZl2p}
\begin{gather}
\nu_{RR} =[Z_y^2\nu_{xx}+Z_x^2\nu_{yy}-2Z_xZ_y\nu_{xy} -G\nu_R \nonumber\\
+(Z_yZ_{xy}-Z_xZ_{yy})\nu_x+(Z_xZ_{xy}-Z_yZ_{xx})\nu_y]/D^{2},\label{DRR}\\
\nu_{ZZ} =[R_y^2\nu_{xx}+R_x^2\nu_{yy}-2R_xR_y\nu_{xy} -F\nu_Z\nonumber\\
+(R_yR_{xy}-R_xR_{yy})\nu_x+(R_xR_{xy}-R_yR_{xx})\nu_y]/D^{2},\label{DZZ}\\
\nu_{RZ} =[(R_xZ_y+R_yZ_x)\nu_{xy}-R_yZ_y\nu_{xx}-R_xZ_x\nu_{yy} \nonumber\\
-G\nu_Z-(Z_yR_{xy}-Z_xR_{yy})\nu_x
-(Z_xR_{xy}-Z_yR_{xx})\nu_y]/D^{2},\label{DRZ}
\end{gather}
\end{subequations}
where we have defined
\begin{gather}
F=R_xD_y-R_yD_x,~G=Z_yD_x-Z_xD_y.
\end{gather}

To calculate second-order derivatives involving $\varphi$, we first obtain the following expression from equation \eqref{xy2RZ}:
\begin{gather}
\nu_\varphi = -R_\zeta \nu_R-Z_\zeta \nu_Z+\nu_\zeta.\label{DP1}
\end{gather}
Making use of equation \eqref{DP1}, we have 
\begin{subequations}\label{Pl2p}
\begin{align}
\nu_{R\varphi}& =-R_\zeta \nu_{RR}-Z_\zeta \nu_{RZ}+\nu_{R\zeta},\label{DRP}\\
%
\nu_{Z\varphi}& =-R_\zeta \nu_{RZ}-Z_\zeta \nu_{ZZ}+\nu_{Z\zeta},\label{DZP}\\
%
\nu_{\varphi\varphi}& =-R_\zeta \nu_{R\varphi}-Z_\zeta \nu_{Z\varphi}+\nu_{\varphi\zeta}.\label{DPP}
\end{align}
\end{subequations}
Note that $\partial_\zeta$ does not commute with $\partial_R$ and $\partial_Z$, so 
\begin{subequations}
\begin{gather}
\nu_{R\zeta} =\nu_{\zeta R}-(D_\zeta/D)\nu_R+(Z_{y\zeta}/D)\nu_x-(Z_{x\zeta}/D)\nu_y,\label{DRz}\\
\nu_{Z\zeta} =\nu_{\zeta Z}-(D_\zeta/D)\nu_Z+(R_{x\zeta}/D)\nu_y-(R_{y\zeta}/D)\nu_x,\label{DZz}\\
\nu_{\varphi\zeta} = -R_\zeta \nu_{R\zeta} -Z_\zeta \nu_{Z\zeta} +\nu_{\zeta\zeta} - R_{\zeta\zeta}\nu_R - Z_{\zeta\zeta}\nu_Z,
\end{gather}
\end{subequations}
where
\begin{subequations}
\begin{gather}
\nu_{\zeta R}=(Z_y/D)\nu_{x\zeta}-(Z_x/D)\nu_{y\zeta},\\
\nu_{\zeta Z} =(R_x/D)\nu_{y\zeta}-(R_y/D)\nu_{x\zeta}.
\end{gather}
\end{subequations}

A subtlety here is that due to the extruded nature of the M3D-$C^1$ elements, mixed second-order derivatives such as $\nu_{x\zeta}$ and $\nu_{y\zeta}$ (but not $\nu_{xy}$) are continuous as well. In the original (tokamak) version, this means that $\nu_{R\varphi}$ and $\nu_{Z\varphi}$ are continuous, such that mixed high-order derivatives like $\nu_{RR\varphi\varphi}$ are allowed and used in equations. However, in non-axisymmetric geometry, $\nu_{R\varphi}$ and $\nu_{Z\varphi}$ are no longer continuous for they also depend on $\nu_{xx}$, etc. That is, we can take strictly no more than second-order physical derivatives on the basis functions in the stellarator version. Hence, some changes such as integration by parts are made to the physics equations to avoid these now-prohibited high-order mixed derivatives.

\section{Treatment of boundary conditions}\label{boundary}
In order to impose boundary conditions, we first need to transform the nodal DoFs from logical $(x,y,\zeta)$ derivatives $(g,g_x,g_y,g_{x},g_{xy},g_{yy},g_\zeta,g_{ x\zeta},g_{ y\zeta},g_{ xx\zeta},g_{ xy\zeta},g_{ yy\zeta})$ to those in terms of the semi-physical $(R,Z,\zeta)$ coordinates, $(g,g_R,g_Z,g_{RR},g_{RZ},g_{ZZ},g_\zeta,g_{ R\zeta},g_{ Z\zeta},g_{ RR\zeta},g_{ RZ\zeta},g_{ ZZ\zeta})$. [For reasons explained in Appendix \ref{second}, it is impossible to transform the logical DoFs into physical $(R,Z,\varphi)$ derivatives.] The inverse transformation is given by
\begin{subequations}\label{p2l}
\begin{gather}
g_x = R_x g_R + Z_x g_Z,\\
g_y = R_y g_R + Z_y g_Z,\\
g_{xx} = R_{xx} g_R + Z_{xx} g_Z + R_x^2g_{RR} \nonumber\\
+Z_x^2g_{ZZ}+2R_xZ_xg_{RZ},\\
g_{xy} = R_{xy} g_R + Z_{xy} g_Z + R_xR_yg_{RR} \nonumber\\
+Z_xZ_yg_{ZZ}+(R_xZ_y+R_yZ_x)g_{RZ},\\
g_{yy} = R_{yy} g_R + Z_{yy} g_Z + R_y^2g_{RR} \nonumber\\
+Z_y^2g_{ZZ}+2R_yZ_yg_{RZ},\\
g_{ x\zeta} = R_x g_{ R\zeta} + Z_x g_{ Z\zeta} + R_{x\zeta} g_R + Z_{x\zeta} g_{Z},\\
g_{ y\zeta} = R_y g_{ R\zeta} + Z_y g_{ Z\zeta} + R_{y\zeta} g_R + Z_{y\zeta} g_{Z},\\
g_{ xx\zeta} =R_{xx} g_{R\zeta} + Z_{xx} g_{Z\zeta} + R_x^2g_{RR\zeta} \nonumber\\
+Z_x^2g_{ZZ\zeta}+2R_xZ_xg_{RZ\zeta}\nonumber\\
+R_{xx\zeta} g_R + Z_{xx\zeta} g_Z + 2R_xR_{x\zeta}g_{RR} \nonumber\\
+2Z_xZ_{x\zeta}g_{ZZ}+2(R_xZ_{x\zeta}+R_{x\zeta}Z_x)g_{RZ},\\
g_{xy\zeta} = R_{xy} g_{R\zeta} + Z_{xy} g_{Z\zeta} + R_xR_yg_{RR\zeta} \nonumber\\
+Z_xZ_yg_{ZZ\zeta}+(R_xZ_y+R_yZ_x)g_{RZ\zeta}\nonumber\\
+R_{xy\zeta} g_R + Z_{xy\zeta} g_Z + (R_{x\zeta}R_y+R_{x}R_{y\zeta})g_{RR}\nonumber\\
 +(Z_{x\zeta}Z_{y}+Z_xZ_{y\zeta})g_{ZZ}\nonumber\\
+(R_{x\zeta} Z_y+R_{x} Z_{y\zeta}+R_{y\zeta}Z_x+R_{y}Z_{x\zeta})g_{RZ},\\
g_{ yy\zeta} =R_{yy} g_{R\zeta} + Z_{yy} g_{Z\zeta} + R_y^2g_{RR\zeta} \nonumber\\
+Z_y^2g_{ZZ\zeta}+2R_yZ_yg_{RZ\zeta}\nonumber\\
+R_{yy\zeta} g_R + Z_{yy\zeta} g_Z + 2R_yR_{y\zeta}g_{RR} \nonumber\\
+2Z_yZ_{y\zeta}g_{ZZ}+2(R_yZ_{y\zeta}+R_{y\zeta}Z_y)g_{RZ}.
\end{gather}
\end{subequations}
Note that $\zeta$ derivatives are always taken after $R$ and $Z$ derivatives here. The direct transformation from logical to semi-physical DoFs is rather cumbersome. In practice, it is more convenient to numerically invert the inverse transformation \eqref{p2l}. 

Now, let us consider a 3D toroidal boundary specified by $R^\mathrm{b}(\theta,\zeta)$ and $Z^\mathrm{b}(\theta,\zeta)$, with the in-plane unit normal and tangential vectors are given by, respectively,
\begin{gather}
\mathbf{n}_\perp = n_1\hat{R}+n_2\hat{Z},~\mathbf{t} = -n_2\hat{R}+n_1\hat{Z},
\end{gather}
where 
\begin{gather}
n_1=Z^\mathrm{b}_\theta/[(R^\mathrm{b}_\theta)^2+(Z^\mathrm{b}_\theta)^2]^{1/2},\nonumber\\
n_2= -R^\mathrm{b}_\theta/[(R^\mathrm{b}_\theta)^2+(Z^\mathrm{b}_\theta)^2]^{1/2}.
\end{gather}
Accordingly, we denote the in-plane normal and tangential derivatives as $g_\mathrm{n}=\mathbf{n}_\perp\cdot\nabla g$ and $g_\mathrm{t}=\mathbf{t}\cdot\nabla g$, respectively. Although $\mathbf{n}_\perp$ is not actually normal to the boundary, in M3D-$C^1$, the magnetic and velocity fields are expressed in terms of a set of scalar fields $(\psi,f,F,U,\omega,\chi)$:
\begin{subequations}
\begin{gather}
\mathbf{B} = \nabla\psi\times\nabla\varphi - \nabla_\perp f_\varphi + F\nabla\varphi,\\
\mathbf{v} = R^2\nabla U\times\nabla\varphi + \omega R^2\nabla\varphi + R^{-2}\nabla_\perp \chi,
\end{gather}
\end{subequations}
such that it is actually the in-plane normal derivative that needs to be constrained in practice.
In the mean time, the other tangential derivative is simply give by $g_\zeta$. Therefore, we can transform the semi-physical DoFs $(g,g_R,g_Z,g_{RR},g_{RZ},g_{ZZ},g_\zeta,g_{ R\zeta},g_{ Z\zeta},g_{ RR\zeta},g_{ RZ\zeta},g_{ ZZ\zeta})$ to $(g,g_\mathrm{n},g_\mathrm{t},g_\mathrm{n n},g_\mathrm{nt},g_\mathrm{tt},g_\zeta,g_{ \mathrm{n}\zeta},g_{ \mathrm{t}\zeta},g_{ \mathrm{n n}\zeta},g_\mathrm{ nt\zeta},g_\mathrm{ tt\zeta})$, i.e., the boundary DoFs. The transformation is given by
\begin{subequations}\label{boundary2}
\begin{gather}
g_\mathrm{n}= n_1g_{R}+n_2g_{Z},\\
g_\mathrm{t}=-n_2g_{R}+n_1g_{Z},\\
g_\mathrm{nn}= n_1^2g_{RR}+2n_1n_2g_{RZ}+n_2^2g_{ZZ},\\
g_\mathrm{nt} = \kappa g_t-n_1n_2g_{RR}+(n_1^2-n_2^2)g_{RZ}+n_1n_2g_{ZZ},\\
g_\mathrm{tt}= -\kappa g_{n}+n_2^2g_{RR}-2n_1n_2g_{RZ}+n_1^2g_{ZZ},\\
g_\mathrm{n\zeta}= n_1g_{R\zeta}+n_2g_{Z\zeta}+\lambda g_\mathrm{t},\label{DNZ}\\
g_\mathrm{t\zeta}=-n_2g_{R\zeta}+n_1g_{Z\zeta}-\lambda g_\mathrm{n},\\
g_\mathrm{nn\zeta}= n_1^2g_{RR\zeta}+2n_1n_2g_{RZ\zeta}+n_2^2g_{ZZ\zeta}\nonumber\\
+2\lambda(g_\mathrm{nt} -\kappa g_\mathrm{t}),\\
g_\mathrm{nt\zeta} =-n_1n_2g_{RR\zeta}+(n_1^2-n_2^2)g_{RZ\zeta}+n_1n_2g_{ZZ\zeta}\nonumber\\
+ \kappa g_\mathrm{t\zeta}+\kappa_\zeta g_\mathrm{t}+\lambda(g_\mathrm{tt}-g_\mathrm{nn}+\kappa g_\mathrm{n}),\\
g_\mathrm{tt\zeta}= n_2^2g_{RR\zeta}-2n_1n_2g_{RZ\zeta}+n_1^2g_{ZZ\zeta}\nonumber\\
-\kappa g_\mathrm{n\zeta}-\kappa_\zeta g_\mathrm{n}-2\lambda(g_\mathrm{nt}-\kappa g_\mathrm{t}),
\end{gather}
\end{subequations}
where $\kappa$ is the curvature of the boundary, and
\begin{subequations}
\begin{gather}
\kappa=(R_\theta Z_{\theta\theta}-Z_\theta R_{\theta\theta})/(R_\theta^2+Z_\theta^2)^{3/2},\\
\lambda=(R_\theta Z_{\theta\zeta}-Z_\theta R_{\theta\zeta})/(R_\theta^2+Z_\theta^2),\\
\kappa_\zeta=(R_{\theta\zeta} Z_{\theta\theta}+R_\theta Z_{\theta\theta\zeta}-Z_{\theta\zeta} R_{\theta\theta}-Z_\theta R_{\theta\theta\zeta})/(R_\theta^2+Z_\theta^2)^{3/2}\nonumber\\
-3\kappa(R_\theta R_{\theta\zeta}+Z_\theta Z_{\theta\zeta})/(R_\theta^2+Z_\theta^2),
\end{gather}
\end{subequations}
where the superscript $\mathrm{b}$ is dropped for convenience. 
Then, we can impose boundary conditions on these boundary DoFs.

\end{document}